\documentclass[aps,pra,showpacs,superscriptaddress,preprint,showkeys]{revtex4}
\usepackage{graphicx}
\usepackage{dcolumn}
\usepackage{bm}
\begin{document}
\title{Constant Roll Inflationary Dynamics with Generalized Potentials in $f(R,\phi,X)$ Gravity}

\author{A. I. Keskin\footnote{aliihsan.keskin@batman.edu.tr; alikeskin039@gmail.com (corresponding author)}}
\affiliation{Department of Electrical and Energy, Besiri OSB
Vocational School, Batman University, Batman, TR-72060, Turkey}
\author{Mehmet Ya\c{s}ar\footnote{mehmetyasar@firat.edu.tr}}
\affiliation{Department of Physics, Faculty of Science, Firat
University, Elazig, TR-23000, Turkey}
\author{K. Kurt\footnote{kadri.kurt@batman.edu.tr; kadridewani@gmail.com }}
\affiliation{Department of Electrical and Energy, Besiri OSB
Vocational School, Batman University, Batman, TR-72060, Turkey}

\begin{abstract}
In this work, we study early-time inflation within a class of $f(R,
\phi, X)$ gravity models under a constant-roll condition. Employing
a generalized potential of the form $V(\phi)^\sigma$, we derive
expressions for the spectral index $n_s$ and tensor-to-scalar ratio
$r$, demonstrating that the inflationary dynamics are primarily
governed by the parameter $\sigma$ and the shape of the potential.
At the upper limit of $n_s$, we obtain a de Sitter-like phase, while
at the lower limit, the model transitions to a quintessence-like
phase through an effective oscillating equation of state parameter
(EoS). Therefore, under the tuning parameter $\sigma < 1$, the model
exhibits a smooth transition from a de Sitter-like phase to a
quintessence-like phase via the oscillating EoS parameter. The
resulting predictions are consistent with recent observations from
the Atacama Cosmology Telescope (ACT), combined with CMB lensing and
DESI BAO data.

\end{abstract}


 \maketitle
\section{Introduction}
One of the central challenges in theoretical cosmology is to
understand the dynamics of the early universe, particularly to
provide an accurate theoretical description of the inflationary
epoch. Inflation was initially modeled using a scalar field, and it
was shown that a near exponential expansion during this era could
resolve several fundamental problems in cosmology, while also
explaining the observed large-scale homogeneity of the universe
\cite{R1, R2, R3, R4}. Observational constraints derived from the
cosmic microwave background (CMB) suggest that the scalar primordial
curvature perturbations are nearly scale-invariant, leading to an
upper bound of approximately $10^{-2}$ on the tensor-to-scalar ratio
$r$ \cite{R5}. Furthermore, the gravitational wave event GW170817
\cite{R6}, resulting from the merger of two neutron stars, confirmed
that the speed of gravitational waves is equal to the speed of light
($c_{T}=1$ in natural units). This observation imposes important
constraints on inflationary models, ruling out or favoring certain
classes of theories. In particular, these constraints have brought
renewed interest in kinetic gravity theories such as the so-called
k-inflation or k-essence models which include non-canonical kinetic
terms and scalar potentials. These frameworks have proven capable of
describing both the inflationary dynamics of the early universe and
the accelerated expansion at late times (i.e., dark energy)
\cite{R7, R8, R9, R10, R11, R12, R13, R14, R15, R16, R17, R18, R19,
R20, R21,
R22}.\\
The present work focuses on the $f(R,\phi,X)$ gravity framework
\cite{R21,R22,R23,R24,R25,R26,R27,R28,R29}, which incorporates the
Ricci scalar $R$, a scalar field $\phi$, and its kinetic term $X =
\frac{1}{2}
\partial^{\mu}\phi \partial_{\mu}\phi$, within a general functional
formalism. In addition to the kinetic contributions, we analyze the
dynamics of a scalar potential under the assumption of a
constant-roll inflationary condition\cite{R30, R31, R32, R33, R34,
R35, R36, R37, R38, R39, R40}. For a broad range of model
parameters, we derive expressions for the scalar spectral index
$n_s$ and the tensor-to-scalar ratio $r$, obtaining results
consistent with current observational data. Moreover, we identify
specific forms of scalar potentials reminiscent of those emerging
from string theory contexts  \cite{R41,R42,R43,R44,R45}.

A key component of our analysis is the equation denoted by $M$,
which involves a geometric formulation of the equation of state
(EoS) parameter, constructed alongside the standard definitions of
pressure and energy density. This equation, which effectively
defines a zero-energy level, leads to an oscillatory behavior of the
EoS parameter through the application of a suitable differential
operator. These oscillations in the EoS parameter induce a
characteristic scaling behavior of the scale factor, driven by the
scalar field. The evolution of the scale factor in this context
depends on both the initial slow-roll parameter and the exponential
potential term $\sigma$. In particular, our analysis of the chaotic
inflationary scenario with $n=2\sigma$ shows that the effective
oscillating EoS parameter can give rise to quintessence-like dark
energy behavior when the condition $\sigma<1$ is satisfied. At the
upper limit of $n_{s}=0.9777$ (coming from recent observations from
the Atacama Cosmology Telescope (ACT)), the model exhibits a de
Sitter-like phase, while approaching the lower limit $n_{s}=0.9709$,
it transitions to a quintessence-like phase. This suggests that the
early universe may have undergone a two-phase evolutionary process,
transitioning from an inflationary de Sitter phase to a quintessence
inflation phase \cite{R47,R48,R49,R50,R51,R52,R53,R54,R55,R56}. We
aim to explore this transition through the lens of modified gravity,
specifically the $f(R,\phi,X)$ formalism, which naturally
accommodates both geometric and scalar degrees of freedom including
kinetic contributions within a unified functional framework. A key
aspect of our investigation involves studying the behavior of the
EoS parameter derived from a geometric construction. We show that
this EoS parameter can exhibit oscillatory behavior, which plays a
crucial role in enabling the transition between cosmological phases.
Moreover, the dynamics of the scalar potential especially for models
with exponential forms $V(\phi)^\sigma$ and $\sigma<1$ are shown to
drive this evolution effectively. By analyzing the $n_{s}$ and $r$
parameters within this framework, and comparing them to current
observational constraints, we aim to demonstrate that such models
not only provide a phenomenologically viable description of the
early universe, but also offer theoretical insights into phase
transitions governed by scalar field dynamics under constant-roll
conditions. In light of these motivations, it is useful to revisit
chaotic inflationary models and explore their phenomenological
generalizations. A particularly simple extension involves modifying
the standard quadratic potential $V(\phi) = \phi^2$ to a generalized
form $V(\phi)^\sigma = \phi^{2\sigma}$, where the exponent $\sigma$
controls the steepness of the potential. When $\sigma < 1$, the
potential becomes flatter, resulting in a prolonged inflationary
phase favored by observations. Conversely, $\sigma > 1$ leads to a
steeper potential and a shorter duration of inflation. This
deformation provides a flexible framework that remains analytically
tractable while capturing a broader range of
inflationary dynamics. \\

\section{The equation of motion and constant-roll condition setup}
We start with the following $f(R,\phi,X)$ gravitational
action~\cite{R21,R22},
\begin{equation}
\int d^{4}x\sqrt{-g}\left[\frac{1}{2}f(R,\phi,X)\right].
\end{equation}
We consider the following functional form for the gravitational
action:
\begin{equation}
f(R,\phi,X)=\frac{R}{\kappa^2} - 2\alpha X - \gamma
V(\phi)^{\sigma},
\end{equation}
where $R$ is the Ricci scalar, $\phi$ is a scalar field, and $X =
\frac{1}{2} \partial^\mu \phi \partial_\mu \phi$ is the kinetic
term. In eq. (2), $\alpha$ is a real parameter; in this study, we
set $\alpha = 1$.  If $\alpha = -1$, it corresponds to a
phantom-type scalar field. Furthermore, $\kappa^{2} =
\frac{1}{M_p^{2}}$ is the reduced Planck mass, and $\gamma$ and
$\sigma$ are free parameters to be analyzed phenomenologically. The
introduction of the exponent $\sigma$ allows us to generalize the
potential and explore its cosmological implications. In the context
of early-universe inflation, introducing a generalized scalar field
potential of the form $V(\phi)^\sigma$ provides a powerful framework
to model a wide range of inflationary dynamics. The exponent
$\sigma$ effectively controls the steepness and curvature of the
inflation potential, which in turn governs the evolution of the
scalar field and the duration of the inflationary epoch. When
$\sigma = 1$, the potential corresponds to the standard form,
commonly used in canonical slow-roll inflation models. However, the
case $\sigma < 1$ allows the field to roll more slowly, and so leads
to an extended inflationary phase, which is often favored by
observational constraints on the scalar spectral index and the
tensor-to-scalar ratio. On the other hand, values of $\sigma
> 1$ steepen the potential, resulting in faster field evolution and
typically a shorter inflationary duration. Such models may be used
to explore departures from conventional slow-roll behavior or to
generate features in the primordial power spectrum. Thus, the
parameter $\sigma$ serves as a tunable degree of freedom that can
interpolate between different inflationary regimes, and may also
emerge naturally from symmetry requirements or effective field
theories. Hence, $\sigma$ provides a phenomenologically rich
parameter that interpolates between different inflationary
scenarios, allowing for compatibility with observational constraints
such as the spectral index $n_s$ and the tensor-to-scalar ratio $r$.

In this study, we adopt the four-dimensional flat
Friedmann-Robertson-Walker (FRW) metric,

\begin{equation}
ds^{2}=-dt^{2}+a(t)^{2}(dx^{2}+dy^{2}+dz^{2}),
\end{equation}
where $a(t)$ is the scale factor of the universe. From the action
(1) with the model (2) we have the following gravitational field
equations, as follows,
\begin{equation}
3H^{2}=\frac{1}{F}(Xf_{x}+\frac{RF-f}{2}-3H\dot{F}),
\end{equation}

\begin{equation}
-2\dot{H}-3H^{2}=\frac{1}{F}(-\frac{RF-f}{2}+\ddot{F}+2H\dot{F}),
\end{equation}

\begin{equation}
(\ddot{\phi}+3H\dot{\phi})f_{x}+\dot{\phi}\dot{X}f_{xx}+f_{\phi}=0.
\end{equation}
Herein, $F=\frac{\partial f}{\partial R}$, $f_{x}=\frac{\partial
f}{\partial X}$ and upper ''dot'' indicates derivative with respect
to cosmic time. Also, for the metric (3) we have
$X=-\frac{\dot{\phi}^{2}}{2}$. Inserting the function (2) into the
field equations we can write
\begin{equation}
3H^{2}=\kappa^{2}(-\alpha X+\frac{\gamma V(\phi)^{\sigma}}{2}),
\end{equation}

\begin{equation}
-2\dot{H}-3H^{2}=\kappa^{2}(-\alpha X-\frac{\gamma
V(\phi)^{\sigma}}{2}),
\end{equation}

\begin{equation}
-2\alpha(\ddot{\phi}+3H\dot{\phi})-\gamma\sigma
V(\phi)^{\sigma-1}V_{\phi}=0.
\end{equation}
Here, in the latest equation the Klein-Gordon equation is obtained
when $\alpha=1$, $\sigma=1$. We shall materialize the inflation
dynamics with the following slow-roll indices \cite{R21,R22,R46},
which have small values,
\begin{equation}
\epsilon_{1}=\frac{\dot{H}}{H^{2}},\,\,\,\,\,\epsilon_{2}=\frac{\ddot{\phi}}{H\dot{\phi}},\,\,\,\,\,
\epsilon_{3}=\frac{\dot{F}}{2HF},\,\,\,\,\,\epsilon_{4}=\frac{\dot{E}}{2HE},
\end{equation}
where
\begin{equation}
E=-\frac{F}{2X}(Xf_{x}+2X^{2}f_{xx}).
\end{equation}
In our case at hand, the eq. (11) equals to
$E=\frac{\alpha}{\kappa^{2}}$. However, the constant-roll condition
has a following general form,
\begin{equation}
\ddot{\phi}=\delta H\dot{\phi}.
\end{equation}
Thus, for model(2), we possess the first and second slow-roll
parameters that characterize the dynamics of inflation. Minimal
values of the initial slow-roll parameters, $\epsilon\ll 1$, are
sufficient to achieve inflation. When the constant-roll parameter is
set to a fixed value, $\epsilon_{2}=\delta$, the inflation of the
universe is determined by the $\epsilon_{1}$ parameter. Observe that
the ultra slow-roll condition arises when $\delta \simeq -3$, which
corresponds to the scenario of a flat potential where $V_{\phi} \sim
0$. However, from the slow-roll inflation viewpoint,
$\frac{\dot{\phi}^{2}}{2}\ll 1$, we obtain

\begin{equation}
H^{2}\simeq \frac{\kappa^{2}\gamma V(\phi)^{\sigma}}{6},
\end{equation}
\begin{equation}
\dot{H}\simeq \frac{\kappa^{2}X\alpha}{2},
\end{equation}

\begin{equation}
\dot{\phi}=-\frac{\gamma\sigma
V(\phi)^{\sigma-1}V_{\phi}}{2\alpha(\delta+3)H}.
\end{equation}
The slow-roll parameters given by (10) is as follows,
\begin{equation}
\epsilon_{1}=-\frac{3\alpha X}{\gamma
V(\phi)^{\sigma}},\,\,\,\,\,\epsilon_{2}=\delta,\,\,\,\,\,
\epsilon_{3}=0,\,\,\,\,\,\epsilon_{4}=0,
\end{equation}

However, the spectral index parameter is as follows\cite{R22},
\begin{equation}
n_{s}=1+2\frac{2\epsilon_{1}-\epsilon_{2}+\epsilon_{3}-\epsilon_{4}}{1+\epsilon_{1}}.
\end{equation}
Also, tensor-to-scalar ratio for theory \cite{R22} is given by
\begin{equation}
r=4(\frac{\Gamma(\frac{3}{2})}{\Gamma(\frac{3}{2}+\epsilon_{2})2^{\epsilon_{2}}}\frac{\dot{\phi}\sqrt{6\alpha}}{\sqrt{\gamma
V^{\sigma}}})^{2}.
\end{equation}
The parameters given by (17) and (18) must be in agreement with the
observation data for inflation to occur. On the other hand, the
amount of inflation is measured by the number of e-foldings $N$,
which is,

\begin{equation}
N=\int_{t_{in}}^{t_{f}}H(t)dt.
\end{equation}
Here $t_{in}$ and $t_{f}$ are the initial and end times of
inflation, and it can be written in terms of a scalar field as
follows\cite{R21},
\begin{equation}
N=\int_{\phi_{in}}^{\phi_{f}}\frac{H}{\dot{\phi}}d\phi.
\end{equation}
The initial inflation field $\phi_{in}$ is defined as a field at
horizon crossing, $\phi_{in} \equiv \phi_{hc}$, as inflation occurs
at that point. This paper examines the exponential representation of
the potential, referred to as chaotic potential\cite{R3}:

\begin{equation}
V(\phi)=V_{0}\phi^{n},
\end{equation}
where $V_{0}=\tau m_{p}^{4-n}$ with $n>0$, and $\tau$ is the
self-coupling constant. With the equations (13), (14) and (15), one
can find the value of the scalar field at the horizon crossing of
perturbations, as follows,
\begin{equation}
\phi_{c}=(\phi_{f}^{2}+\frac{2\sigma
nN}{\alpha(\delta+3)\kappa^{2}})^{\frac{1}{2}}.
\end{equation}
The scalar field value at the end of inflation should also be
established. For this one can use $\epsilon_{1}(\phi_{f})\simeq 1$
\cite{R21,R41}, indicating the value of the slow-roll parameter at
the end of inflation. Consequently, we obtain

\begin{equation}
\phi_{f}=(-\frac{9n^{2}\sigma^{2}}{4\kappa^{2}\alpha(\delta+3)^{2}})^{\frac{1}{2}},
\end{equation}
and the resulting slow-roll parameters is calculated as follows,
\begin{equation}
\epsilon_{1}=\frac{-9n\sigma}{-9n\sigma+8N(\delta+3)},\,\,\,\,\epsilon_{2}=\delta,\,\,\,\,\epsilon_{3}=\epsilon_{4}=0.
\end{equation}
It is observed that inflation depends on the first two slow-roll
parameters. The second slow-roll parameter will be fixed within the
observation data frame, thus appearing to a two-stage inflation
model.

\section{A Generalization of a Chaotic Potential and Comparison with Observational Data}
Starting from the well-known chaotic inflation model with a
quadratic potential ($n = 2$),
\begin{equation}
V(\phi) = \phi^{2},
\end{equation}
one can generalize the potential by introducing $\sigma$ as
\begin{equation}
V(\phi)^\sigma = \phi^{2\sigma}.
\end{equation}
For example, the case $\sigma = 1$ recovers the standard quadratic
potential, while $\sigma = 2$ yields the quartic form $V(\phi) =
\phi^4$. Specific choices, such as $\sigma = \frac{2}{5}$ or $\sigma
= \frac{1}{3}$, lead to potentials of the form $V(\phi) \propto
\phi^{\frac{4}{5}}$ and $V(\phi) \propto \phi^{\frac{2}{3}}$,
respectively, which are motivated by constructions in certain
string-inspired models.

This generalization leads to a deformation of the standard scenario
that can have significant implications for inflationary dynamics. It
allows the model to interpolate between different inflationary
behaviors while preserving the simplicity of monomial potentials. In
this framework, the exponent $\sigma$ effectively modulates the
steepness of the potential, thereby controlling the evolution of the
inflation field. Moreover, such a generalized form may help
reconcile chaotic inflation with current observational bounds on the
scalar spectral index $n_s$ and the tensor-to-scalar ratio $r$,
which tend to disfavor the standard quadratic potential. From a
theoretical perspective, the exponent $\sigma$ could arise from
high-energy corrections, symmetry considerations, or effective field
theory expansions, making this generalization both
phenomenologically relevant and theoretically motivated.

On the other hand, recent ACT measurements \cite{R58,R59} suggest
slightly higher values of $ n_s$ compared to earlier Planck
results\cite{R5}. ACT data combined with Planck and BAO data show a
constraint as follows,

\begin{equation}
n_{s} = 0.9743 \pm 0.0034, \quad r < 0.038 \quad \text{(95\% C.L.,
ACT-DR6 + Planck + BAO)}.
\end{equation}
Based on this data, the inflationary parameters are analyzed. By
employing equations (17), (18) and (24), we derive the expressions
for the spectral index and tensor-to-scalar ratio as follows:
\begin{equation}
n_{s}=\frac{9n\sigma(-3+\delta)+4N(\delta+3)(1-2\delta)}{-9n\sigma+4N(\delta+3)},
\end{equation}

\begin{equation}
r=4[\frac{-6n\sigma\Gamma(\frac{3}{2})}{\Gamma(\frac{3}{2}+\delta)2^{\delta}\sqrt{-9n^{^{2}}\sigma^{2}+8\sigma
nN(\delta+3)}}]^{2}.
\end{equation}
The spectral index parameter and the tensor-to-scalar are dependent
upon three factors, $n,\sigma,\delta$, and are independent of
additional parameters, $V_{0}, \gamma$. From (28) and (29), we can
derive the relationship between $n_{s}$ and $r$.

\begin{equation}
n_{s}=\frac{4^{1-\delta}\pi-2^{3-2\delta}\pi\delta-17r\Gamma(\frac{3}{2}+\delta)^{2}
+4r\delta\Gamma(\frac{3}{2}+\delta)^{2}}{4^{1-\delta}\pi-5r\Gamma(\frac{3}{2}+\delta)^{2}}.
\end{equation}

Using the equality given in (30), we obtain the lower bound value of
$n_{s} = 0.9709$ when the tensor-to-scalar ratio takes its upper
limit of $r = 0.038$ and $\delta = 0.0002829$. For a value of the
tensor-to-scalar ratio close to its minimum, $r = 0.00013$, with
$\delta = 0.0111$, we obtain the upper bound value of $n_{s} =
0.9777$. The values $n_{s} = 0.9709$ and $n_{s} = 0.9777$ correspond
to the lower and upper bounds, respectively, obtained from eq.(30).
In this analysis, no parameter has been fixed; the results are
obtained by allowing $r$ and $\delta$ to vary freely within their
physically meaningful ranges. As a result, the corresponding values
of $\delta$ lie within the interval $0.0002829 \leq \delta \leq
0.0111$. In the context of eq.(30), changes in $\delta$ affect the
scalar spectral index $n_s$, which is closely related to the
dynamics of inflation. Therefore, $\delta$ serves as an effective
parameter in regulating the inflationary behavior of the model.

On the other hand, taking $\delta = 0.0002829$ with $n_{s} =
0.9709$, the product $n\sigma$ can be obtained from eq.(28) as
$n\sigma \cong 1.12526$. Similarly, for the upper limits of $\delta
= 0.0111$ and $n_{s} = 0.9777$, we obtain $n\sigma \cong 0.00399$.
Hence, for the upper and lower limits of $n_{s}$ and $r$ we have
\begin{equation}
0.00399<n\sigma<1.12526.
\end{equation}
Since we examine a generalized form of the chaotic potential given
by (26), the effect of the parameter $\sigma$, introduced
phenomenologically, can be discussed within this framework. for
$n=2$ we have

\begin{equation}
0.00199622<\sigma<0.6263.
\end{equation}

\begin{figure}[ht]
\centering
\begin{minipage}{0.48\textwidth}
  \centering
  \includegraphics[width=\linewidth]{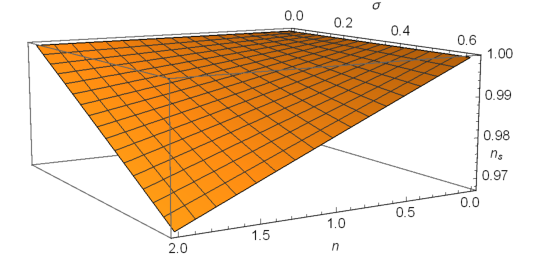}
\end{minipage}%
\hfill
\begin{minipage}{0.48\textwidth}
  \centering
  \includegraphics[width=\linewidth]{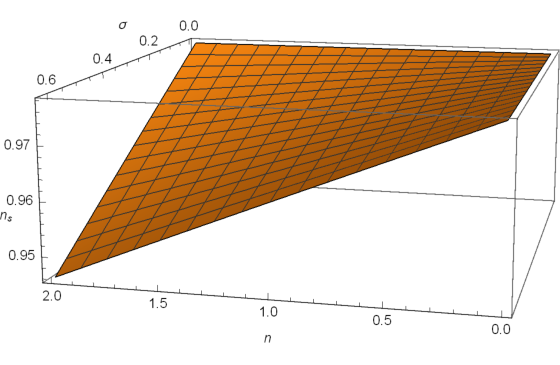}
\end{minipage}
\caption{Variation of the scalar spectral index $n_{s}$ as a
function of $n \in (0,2)$ and $\sigma \in (0.00199622, 0.6263)$ for
two fixed values of the parameter $\delta$. The left panel
corresponds to $\delta =0.0002829$, where higher values of $\sigma$
are preferred for $n = 2$, resulting in $n_{s}$ values near the
lower end of the observational range. The right panel corresponds to
$\delta = 0.0111$, where lower values of $\sigma$ are favored,
leading to $n_{s}$ values close to the upper observational limit.
Both cases remain fully consistent with the current bounds from ACT
data.} \label{fig:ns_dual}
\end{figure}
The considered range of the parameter $\sigma \in (0.00199622,
0.6263)$ not only satisfies current observational bounds on the
$n_{s}$ but also carries theoretical significance. In particular,
this interval has been explored for different values of the
parameter $\delta$, such as $\delta = 0.0002829$ and $\delta =
0.0111$, which correspond respectively to lower and upper limits on
$n_{s}$ within the observationally allowed range (see FIG. 1). For
small values of $\sigma$ (e.g., $\sigma \sim 0.00199622$), the
potential becomes relatively flat, supporting a slow-roll evolution
of the scalar field and yielding an effective EoS close to
$w_{e}\sim -1$. This correspondence shows that the model space
analyzed here is broad enough to interpolate between near de Sitter
inflation and slowly evolving scalar field dynamics. As will be
discussed under a dedicated section below, such slow-roll behavior
can also be associated with quintessence-type inflationary models,
which further motivates the use of this parameter space.

\subsection{Quintessence transition surrounding the de Sitter phase}

In the classical cosmological inflationary framework (where $\sigma
= 1$), the inflationary field evolves on a high and flat potential
under the slow-roll approximation, resulting in an exponential
growth of the scale factor. However, while the point $w = -1$
corresponds to a de Sitter phase in the context of late-time
acceleration, this accelerated expansion during the early universe
is characterized by a quasi-de Sitter phase ($w \sim -1$). The
effective EoS parameter is defined geometrically as follows:
\begin{equation}
w_{e} = -1 - \frac{2 \dot{H}}{3 H^{2}}.
\end{equation}
Conversely, the dynamical characteristics of dark energy are
described by quintessence, which corresponds to
\begin{equation}
-1 < w_{e} < -\frac{1}{3}.
\end{equation}
Specifically, regarding the nature of late-time dark energy, some
models suggest that the quintessence field is confined within extra
dimensions \cite{R51,R53} during the early epoch of the universe.
Consequently, the notion of "quintessence," a dynamic form of dark
energy, may have originated in the primordial universe. In this
light, we investigate the presence of a quintessence regime within
the proposed model during the early inflationary period. Moreover,
in the context of string theory, it has been shown that scalar
fields emerging from moduli or axionic sectors of compactified extra
dimensions can exhibit quintessence-like behavior
\cite{R60,R61,R62,R63,R64}. Such fields are favored by the Swampland
conjectures, which suggest that stable de Sitter vacua are difficult
to realize in a UV-complete theory, while slowly varying potentials
as in quintessence models remain viable. Therefore, exploring a
quintessence phase in the early universe is not only
phenomenologically relevant, but also theoretically well motivated.
The present model, characterized by a potential of the form
$V(\phi)^{\sigma}$, naturally accommodates both a de Sitter-like
inflationary regime and a subsequent quintessence-like phase,
providing a smooth and consistent transition between these cosmic
eras. To analyze this transition, we start with the Friedmann
equations (7) and (8), which can be written as follows:

\begin{equation}
\rho_{e} = (-\alpha X + \frac{\gamma V(\phi)^{\sigma}}{2}), \quad
p_{e} = (-\alpha X - \frac{\gamma V(\phi)^{\sigma}}{2}).
\end{equation}
The resulting effective EoS parameter is given by
\begin{equation}
w_{e} = \frac{p_{e}}{\rho_{e}}.
\end{equation}
From (33) and (36) we can write the expression\cite{R53,R54,R56},
\begin{equation}
M = [p_{e} - w \rho_{e} - 2 \dot{H} - 3 H^{2} (1 + w_{e})].
\end{equation}
In this expression matter and geometry are in a unified form if
$M=0$. To transform this expression into linear form because of the
exponential term ($\sigma$), we will employ the following
differential operator \cite{R48,R54,R55,R56}:
\begin{equation}
D = \frac{d}{dt}\left(\frac{\phi}{\ddot{\phi}} \frac{d}{dt}\right).
\end{equation}
This operator can be interpreted as a time-dependent, weighted
second-order derivative modulated by the scalar field and its
acceleration, effectively capturing deviations from standard
slow-roll dynamics. By using the definition of the effective EoS
parameter, we derive the subsequent equality:
\begin{equation}
\phi \ddot{\phi} + \dot{\phi}^{2} = \frac{-\gamma \sigma n^{2}
V(\phi)^{\sigma}}{\alpha \delta} \left[\frac{\dot{\phi} \sigma}{\phi
H} - \epsilon_{1}\right].
\end{equation}
Here, the oscillations of the effective EoS parameter necessitate
the requirement
\begin{equation}
\phi \ddot{\phi} + \dot{\phi}^{2} = 0,
\end{equation}
which therefore demands that
\begin{equation}
\left[\frac{\dot{\phi} \sigma}{\phi H} - \epsilon_{1}\right] = 0.
\end{equation}
This naturally drives the right-hand side of the equation to zero as
well, yielding a dynamical condition that links $\phi$,
$\dot{\phi}$, $\sigma$, the Hubble parameter $H$, and the first
slow-roll parameter $\epsilon_1$ as specified in (24). While the
left-hand side captures the kinetic signature of an oscillating
scalar field, the right-hand side encodes a scaling relation that
characterizes the behavior of the scale factor in terms of $\phi$.
Therefore, this equation bridges the oscillatory dynamics of the
scalar field with the background expansion governed by the scalar
potential shape $V(\phi)^\sigma$. From the eq. (40), we derive
\begin{equation}
\dot{\phi}^{2} = \left(\frac{\gamma \sigma n}{2 \alpha}\right)
V(\phi)^{\sigma},
\end{equation}
leading to the subsequent oscillatory effective EoS parameter
\cite{R48,R49,R54,R57}:
\begin{equation}
w_{e} = \frac{n \sigma - 2}{n \sigma + 2}.
\end{equation}
From right side of (39) or from (41) we obtain the scale factor in
terms of scalar field,
\begin{equation}
a = a_{0} \phi^{\frac{\sigma}{|\epsilon_{1}|}},
\end{equation}
where $a_{0}$ is an integral constant. The scale factor is dependent
upon $\sigma$ and the parameter $\epsilon_{1}$, and it exhibits a
scaling behavior with respect to the scalar field $\phi$. However,
the first slow-roll parameter are computed as follows:
\begin{equation}
0.000024951<\mid\epsilon_{1}\mid<0.00788977,
\end{equation}
where we used the range (32) with the lower bound of
$\gamma=0.0002829$. While the precise initial conditions of
inflation remain uncertain, an adjustment can be implemented using
the integral constant in the scale factor. As $a_{0}$ approaches
zero, an initial situation of nearly a singular case can be
achieved. When the inflationary field emerges at a quantum scale, we
can discuss fast-accelerated expansion, characterized by the range
(32) and a chaotic potential represented by $n=2$. Furthermore, for
$n=2$, the range (32) yields the following inequality in the
effective EoS:
\begin{equation}
-0.996016< w_{e} <-0.229785,
\end{equation}
which shows a quintessence phase. Besides this, for the quartic
potential form $V(\phi) = \phi^{4}$, the range $0.0009975 < \sigma <
0.31315$ is obtained. Therefore, the same range as in eq. (46) is
recovered. Hence, one can calculate the same effective EoS values
for different values of $n$. This indicates that the $\sigma$
parameter acts as a tuning parameter to eliminate potential
steepening that may occur in the flat regions of the potential at
high $n$ values, thereby playing a crucial role in controlling the
inflationary field. The range of $\sigma$ values obtained from the
ACT data for the $n_{s}$ and $r$ keeps the shape of the potential
function within the range of (46). While a de Sitter-like phase is
observed at the lower bound of $\sigma\cong 0.00199622$, where
$n_{s} = 0.9777$, a transition toward a quintessence phase appears
as the value approaches the upper bound of $\sigma\cong 0.6263$
where $n_{s}= 0.9709$.

\section{Closing remarks}

Understanding the dynamics of the early universe, especially the
inflationary epoch, remains one of the most profound challenges in
theoretical cosmology. Accurately modeling this phase is crucial not
only for explaining the large-scale structure and homogeneity of the
universe but also for shedding light on the nature of fundamental
fields that drove inflation. Traditional inflationary models based
on canonical scalar fields have successfully addressed many issues
but often lack the flexibility to naturally describe phases
immediately following the initial inflationary stage. In this
context, phenomenological extensions such as the $f(R,\phi,X)$
gravity model, incorporating curvature, scalar fields, and kinetic
terms with additional parameters like $\sigma$, $\delta$, and $n$,
offer a promising framework. These extensions provide richer
dynamics, enabling us to explore scenarios where the universe
transitions from an initial de Sitter phase to a quintessence-type
inflationary phase before reheating.

This paper discusses a two-stage inflation scenario within the
framework of the phenomenological gravity model $f(R,\phi, X)$.
Three parameters, namely $\sigma$, $\delta$, and $n$, significantly
influence the inflationary dynamics. The interval derived in (32) by
using ACT data is particularly important for the potential form
$V(\phi)^\sigma$. Focusing on the case $n=2$, which represents a
generalization of a chaotic potential given by (26), we analyzed the
vacuum-like equation $M=0$, encompassing the EoS parameter. By
applying a differential operator designed to linearize the
exponential term in the $V(\phi)^\sigma$ potential within the
$f(R,\phi,X)$ framework, we derived expressions describing the
scalar field evolution alongside an oscillating EoS parameter (43).
Our results, especially the equality in (43), show that the
combination of $n=2$ and the range in (32) leads to
quintessence-type oscillations. This suggests that the universe may
undergo a transition from an initial de Sitter (slow-roll) vacuum
phase to a quintessence-like inflationary regime occurring
immediately after, prior to reheating. Furthermore, the scale factor
exhibits a characteristic scaling behavior, which is influenced not
only by the first slow-roll parameter $\epsilon_1$ but also by the
potential exponent $\sigma$. These findings emphasize the critical
role of $\sigma$ as a tuning mechanism that controls the
inflationary dynamics. In particular, it helps to suppress the
steepening of the potential that may arise in flatter regions at
higher $n$ values. The results are obtained in light of the recently
published observational data from ACT. The range of $\sigma$ values
constrained by ACT observational data for the spectral index $n_s$
and the tensor-to-scalar ratio $r$ ensures that the potential
remains within the physically viable bounds indicated by (46).
Notably, the lower limit of $\sigma \approx 0.00199622$,
corresponding to $n_s = 0.9777$, is associated with a de Sitter-like
inflationary phase. As $\sigma$ increases toward its upper bound
$\approx 0.6263$, where $n_s = 0.9709$, the system undergoes a
smooth transition into a quintessence regime. It has been identified
that when inflation starts near the upper limit of the spectral
index $n_s$ and evolves toward its lower bound, a transition into a
quintessence phase emerges a behavior that is governed by the tuning
parameter $\sigma$.

In addition, the generalization of the chaotic potential through the
introduction of the exponent $\sigma$ provides a flexible and
theoretically motivated extension of the classical quadratic
potential. This modification allows the inflationary model to better
accommodate observational constraints on the $n_s$ and $r$, offering
a smoother transition between inflationary phases. In other words,
within the interval given by (31), each value of $n$ constrained by
the upper and lower bounds of $n_s$ and $r$ from the ACT
observational data is forced-through the tuning parameter $\sigma$
to remain within the inflationary phase defined by the same
interval. This, in turn, places the effective oscillatory EoS
parameter given by (43) within the bounds specified in (46).
Consequently, the phenomenological parameter $\sigma$ effectively
emerges as a control or tuning parameter governing the inflationary
dynamics under constant-roll inflation condition.

Although quintessence is typically studied in the context of
late-time cosmic acceleration, its emergence in the early universe
particularly immediately following a de Sitter-like phase carries
significant theoretical interest. In the context of string theory
and extra dimensional models, slowly rolling scalar fields with
nontrivial potentials are more consistent with high energy
UV-complete theories than strictly flat de Sitter vacua. The
appearance of a quintessence-type phase in this model, regulated by
the tuning parameter $\sigma$, therefore not only enables a smooth
inflationary evolution but also aligns with fundamental theoretical
frameworks. This enhances the physical relevance of the model by
linking early-time inflationary dynamics with scalar field behaviors
anticipated in fundamental physics.

\end{document}